\documentclass[prl,graphicx,amsmath,amssymb,reprint]{revtex4-1}

\usepackage{graphicx}

\draft 

\begin{document}


\title{A proper scalar product for tachyon representations in configuration space} 



\author{F.F. L\'opez-Ruiz}
\email{paco.lopezruiz@uca.es}
 \affiliation{Departamento de F\'\i sica Aplicada, Universidad de C\'adiz, Campus de Puerto Real, E-11510 Puerto Real, C\'adiz, Spain}
\author{J. Guerrero}%
 \email{julio.guerrero@ujaen.es}
\affiliation{ 
Departamento de Matem\'aticas, Universidad de Ja\'en, Campus las Lagunillas, 23071 Ja\'en, Spain
}%
\altaffiliation[Also at ]{Institute Carlos I of Theoretical and Computational Physics, University of Granada, Fuentenueva s/n, 18071 Granada, Spain}

\author{V. Aldaya}
\email{valdaya@iaa.es}
\affiliation{%
Instituto de Astrof\'\i sica de Andaluc\'\i a (IAA-CSIC), Glorieta de la Astronom\'\i a, E-18080 Granada, Spain
}%


\date{\today}

\begin{abstract}
We propose a new inner product for scalar fields that are solutions of the Klein-Gordon  equation with $m^2<0$. This inner product is non-local, bearing an integral kernel including Bessel functions of the second kind, and the associated norm proves to be positive definite in the subspace of oscillatory solutions, as opposed to the conventional one. Poincar\'e transformations are unitarily implemented on this subspace, which is the support of a unitary and irreducible representation of the proper orthochronous Poincar\'e group. We also provide a new Fourier Transform between configuration and momentum spaces which is unitary, and recover the projection onto the representation space. This new scenario suggests a revision of the corresponding quantum field theory. 
\end{abstract}

\pacs{Valid PACS appear here}

\maketitle 

\textit{Introduction.---} The Klein-Gordon equation with negative squared mass, $m^2= -\kappa^2<0$, generated an interesting debate in the late 60's and the 70's. Assuming the usual signature $(+,-,-,-)$ as well as units in which $\hbar=c=1$, it is written
\begin{equation}
    \partial_\mu \partial^\mu \phi - \kappa^2 \phi = 0 \,.
    \label{KGtacheq}
\end{equation}
Originally, the corresponding quantum field theory was thought to represent 
faster-than-light particles \cite{Feinberg1}. However, 
the theory did not succeed, one of the main issues being the problems with the causality \cite{Aharonov} and 
the unitary implementation of Poincar\'e transformations on this simple model \cite{Sudarshan1,Sudashan2,Schroer,Jue,Feinberg2}. In this sense, although
the representations of the Poincar\'e group for $m^2<0$ were not originally analyzed by Wigner 
\cite{Wigner}, later authors constructed such representations, but only in momentum
 space \cite{Moses,Fonda,Mukunda} with support on the one-sheeted hyperboloid. A unitary and
 irreducible representation realized on configuration space was still, surprisingly, lacking. 

More recently, tachyonic modes were found to be present in many different physical models \cite{Sen,Armoni,ellis,Landulfo,Paddy,Teixeira,Felder1,Felder2,Yokota,Oriekhov,Trachenko}, and have been assumed to represent unstable states.  Regardless of the nature of tachyonic modes, 
either fundamental or not, or their actual role in any physical theory, our aim in this 
Letter is to point out that a consistent description of scalar solutions of \eqref{KGtacheq} cannot rely on the standard inner product
%
%
\begin{equation}
\label{KGinnpro}
\langle \phi,\varphi\rangle_{st} = 
i \int_\Sigma d\sigma^\mu 
\Big(\phi(x)^*\partial_\mu\varphi(x) -
\partial_\mu\phi(x)^*\varphi(x)\Big)\,,
\end{equation}

\noindent where $x=(x^0,\vec x)$, $\Sigma$ is a spatial Cauchy hypersurface and $d\sigma^\mu$ 
its surface element, but rather on the proposed inner product
\begin{widetext}
\begin{equation}
\label{KGtachinnprocov}
\langle \phi,\varphi\rangle = \frac{\kappa^2}{4\pi} 
\int_\Sigma d\sigma^\mu  \int_{\Sigma'} d\sigma^{\nu'}  
\Big( \phi(x)^*  
\partial_\mu\partial_{\nu'} K(x-x')
\varphi(x')  
+\partial_\mu \phi(x)^* 
K(x-x') 
\partial_{\nu'} \varphi(x')  \Big)\,,
\end{equation}
where $K(x)=-\frac{Y_1(\kappa s)}{\kappa s}$ with $s=\sqrt{\vec x\,^2-(x^0)^2}$, and $Y_1(x)$
is a Bessel function of the second kind. If we choose to fix $\Sigma$ at $x^0=0$, we obtain the
more convenient form

\begin{equation}
\label{KGtachinnpro}
\langle \phi,\varphi\rangle = \frac{\kappa^2}{4\pi} \int d^3x \int d^3x' 
\Big( \phi(x)^*  
k_2(|\vec x-\vec x{\,}'|)
\varphi(x')  
+\partial_0 \phi(x)^* 
k_1(|\vec x-\vec x{\,}'|) 
\partial_{0'} \varphi(x')  \Big)\,,
\end{equation}
\end{widetext}
where $k_2(r)=\frac{\tilde{Y}_2(\kappa r)}{r^2} $ and $k_1(r)=-\frac{Y_1(\kappa r)}{\kappa r}$. 
The tilde in $\tilde{Y}_2(x)$ indicates that a proper regularization is required (see later). 
Note that, even though the kernels $k_2(r)$ and $k_1(r)$ are oscillatory, they are positive-definite kernels on a suitable subspace (see below). With the aid of \eqref{KGtachinnprocov}-\eqref{KGtachinnpro}, we want to show that the quantum theory of a scalar free `tachyon' field with wave equation \eqref{KGtacheq} is self-consistent and behaves as the scalar $m^2<0$ representation of the Poincar\'e group. 

\textit{The space of oscillatory solutions.---} We define the Hilbert space of the theory $\mathcal H_{osc}$ as the closure of the linear span of solutions of \eqref{KGtacheq} made of exponentials with real frequencies, that is, 

\begin{equation}
\label{planewaves}
\psi_{\vec p \pm}(x)= \frac{1}{\sqrt{2(2\pi)^3}} e^{-i p_\mu x^\mu}\!,\quad p_0 = 
\pm \omega_p \equiv \pm \sqrt{\vec{p}\,^2-\kappa^2}\,,
\end{equation}
where $|\vec p\,|\ge \kappa$ necessarily. Therefore, there is a gap in the norm of spatial momentum. This subspace of solutions will be called oscillatory solutions, since they represent solutions with oscillatory behavior, in contrast with solutions with $|\vec p\,|<\kappa$ for which $p_0$ is imaginary and therefore they grow or decrease exponentially in time. The gap in spatial momentum also introduces a spatial oscillatory behavior. Note that this classification is Lorentz (and Poincar\'e) invariant, since a boost 
$p_\mu\rightarrow p'_\mu$ does not change the relation 
${p'}_0^2-|\vec p \,'|^2+\kappa^2=0 \Rightarrow |\vec p \,'|^2=\kappa^2+{p'}_0^2\geq \kappa^2$.

The restriction to $|\vec p\,|\ge \kappa$ had already been noted by Feinberg in Ref.~\onlinecite{Feinberg1}. Now, mainly in light of \eqref{KGtachinnpro}, we highlight several properties singularizing that space. First, the initial value problem defined by
 \eqref{KGtacheq}, $\Sigma$ and prescribed initial values $\phi(\vec x)=\phi(x)|_\Sigma$ and $\dot \phi (\vec x)= \partial_0\phi(x)|_\Sigma$, is well-posed for  oscillatory solutions, i.e., the
  solution's behavior changes continuously with \textit{properly} given initial conditions, whereas is ill-posed for real exponential
  solutions. We put emphasis on `properly' because  $\phi(\vec x)$ and $\dot\phi(\vec x)$ cannot be arbitrary functions belonging to $L^2(\Sigma)$: they must satisfy that their Fourier transforms $\hat\phi(\vec p\,)$ and $\hat{\dot \phi}(\vec p\,)$ have support on $|\vec p\,|\in [\kappa,+\infty )$. 

Second, Poincar\'e infinitesimal generators $P_\mu=i\partial_\mu$ and 
$M_{\mu\nu}= x_\mu P_\nu-x_\nu P_\mu$ are Hermitian with respect to the inner product \eqref{KGtachinnpro}.
In order to check this explicitly, some recurrence relations for the Bessel functions are 
required (see Ref.~\onlinecite{S3mom} for a similar computation). The inner product proves
 to be invariant under Poincar\'e transformations and, in 
particular, time-invariant (the proof of this statement is parallel to that presented in Ref.~\onlinecite{Wolf} for solutions of the Helmholtz equation). This way, $\phi(x)$ and $\partial_0 \phi(x)$ in
 \eqref{KGtachinnpro} can be replaced by initial conditions on $\Sigma$,
$\phi(\vec x)$ and $\dot\phi(\vec x)$. 
  
  Also, the representation of the proper orthochronous Poincar\'e  group $\mathcal P_+^{\uparrow}$ on oscillatory solutions with the inner product \eqref{KGtachinnpro}, given by the exponentiation of $P_\mu$   and $M_{\mu\nu}$, is irreducible. That includes both positive and negative  frequencies, in contrast with the $m^2>0$ case. A way to verify this is by introducing a unitary transformation (see Eqs. \eqref{FT}-\eqref{IFT}) between the space of oscillatory solutions and the momentum space representation, where the unitary and irreducible representation for $m^2<0$ is known \cite{Fonda}.
  
  Finally, the inner product \eqref{KGtachinnpro} turns out to define a positive definite norm on the space of oscillatory solutions, which is key to establish the equivalence between the configuration space representation and the known momentum space representation, with support on the one-sheet hyperboloid and which bears a positive definite inner product: the well-known expression \eqref{KGinnpro} needs to be discarded for the configuration space representation as it fails to be positive definite. 

\textit{Unitary equivalence with momentum space.---} In the momentum space realization of the scalar representation \cite{Fonda}, the Hilbert space $\mathcal H_m$ is given by functions on the one-sheeted hyperboloid normalizable with respect to the integration measure $d^4\mu(p) \propto \delta(p^2+\kappa^2)d^4p$. The Cartesian projection on $p_0=0$ requires distinguishing a positive frequency component $\phi^+(\vec p\,)$ and a negative frequency one $\phi^-(\vec p\,)$, in such a way that $\phi(\vec p\,)=(\phi^+(\vec p\,),\phi^-(\vec p\,))$. The inner product is then given by
\begin{equation}
	\label{KGtachinnpromom}
	\langle \phi,\varphi\rangle_m = \int_{|\vec p\,|>\kappa}\!\frac{d^3p}{\omega_p}
	\Big(\phi^+(\vec p\,)^*\varphi^+(\vec p\,)+\phi^-(\vec p\,)^*\varphi^-(\vec p\,)\Big)\;.
\end{equation}
The subspaces of positive and negative frequencies are orthogonal, but they are not invariant subspaces since Lorentz boosts mix them up. Note also that this inner product is non-degenerate on functions with support on $|\vec p\,|>\kappa$, in agreement with $\mathcal H_m$. 

The momentum space counterpart of \eqref{planewaves}, eigenstates of momentum $\vec p\,'$, are given by (recall that $|\vec p\,'|\geq \kappa$ is required)  $\psi_{\vec p\,'+}(\vec p\,)=(\omega_p \delta^{(3)}(\vec p-\vec p\,'),0)$ (for positive frequency states), and $\psi_{\vec p\,'-}(\vec p\,)=(0,\omega_p \delta^{(3)}(\vec p-\vec p\,'))$ (for negative frequency states). The scalar product among them gives 
\begin{equation}
\label{A15}
	\langle \psi_{\vec p \sigma},\psi_{\vec p\,'\sigma'}\rangle_m =  \delta_{\sigma\sigma'} \omega_p \delta^{(3)}(\vec p-\vec p\,')\,,
\end{equation}
where $\sigma,\sigma'=\pm$. 

Let us find the transformation taking momentum space wavefunctions 
$(\phi^+(\vec p\,),\phi^-(\vec p\,))$ to configuration space wavefunctions characterized by initial conditions $\phi(\vec x)$ and $\dot\phi(\vec x)$. Any oscillatory solution to \eqref{KGtacheq} can be expanded in terms of \eqref{planewaves}: 
\[
\phi(x) = \int_{|\vec p\,|>\kappa}\!\frac{d^3p}{\omega_p}
	\Big(\phi^+(\vec p\,)\,\psi_{\vec p +}(x)+\phi^-(\vec p\,)\,\psi_{\vec p -}(x)\Big)\;.
\]
Differentiating with respect to $x^0$ and making $x^0=0$ gives: 
\begin{align}
\label{FT}
\phi(\vec x) &= \frac{1}{\sqrt{2(2\pi)^3}}\int_{|\vec p\,|>\kappa}\!\frac{d^3p}{\omega_p}
	\Big(\phi^+(\vec p\,)+\phi^-(\vec p\,)\Big)e^{i\vec p \cdot \vec x}\;,
	\\
\dot \phi(\vec x) &=\frac{-i}{\sqrt{2(2\pi)^3}} \int_{|\vec p\,|>\kappa}\!\!\!\!d^3p
	\Big(\phi^+(\vec p\,)-\phi^-(\vec p\,)\Big)e^{i\vec p \cdot \vec x}\;.
\label{FTd}
\end{align}

Let us now define the following inverse transformation: 
\begin{equation}
\label{IFT}
	\phi^\pm(\vec p\,)=\!\frac{1}{\sqrt{2(2\pi)^3}}\!\!
	\int \!\!d^3x \Big(\! (\omega_p)_+ \phi(\vec x)\pm 
	i (\omega_p)_+^0 \dot\phi(\vec x)\!\Big)e^{-i\vec p \cdot \vec x}\,,
\end{equation}
where $(\omega_p)_+ = \omega_p$ for $|\vec p\,|\geq \kappa$, $(\omega_p)_+ = 0$ for $|\vec p\,|< \kappa$, $(\omega_p)_+^0 = 1$ for $|\vec p\,|\geq \kappa$ and $(\omega_p)_+^0 = 0$ for $|\vec p\,|< \kappa$. Eq. \eqref{IFT} together with \eqref{FT} and \eqref{FTd} define a 
`Fourier' transform and its inverse, relating configuration and momentum spaces. In particular, it can be checked that eigenstates of momentum $\vec p\,'$ in configuration space are mapped into eigenstates of momentum $\vec p\,'$ in momentum space and vice versa. 

We can now show that both configuration and momentum space representations on $\mathcal H_{osc}$ and $\mathcal H_m$ respectively are unitarily equivalent. Given that $\psi_{\vec p\sigma}(\vec x)=\frac{1}{\sqrt{2(2\pi)^3}}e^{i\vec p \cdot \vec x}$ and $\dot \psi_{\vec p\sigma}(\vec x)=\frac{-i\sigma \omega_p}{\sqrt{2(2\pi)^3}}e^{i\vec p \cdot \vec x}$, with $\sigma=\pm$, let us compute \eqref{A15} in configuration space using \eqref{KGtachinnpro} (we insert $e^{i\vec p \cdot \vec x'}e^{-i\vec p \cdot \vec x'}$ when necessary):

\begin{widetext}
	\begin{align}
&\langle \psi_{\vec p\sigma} , \psi_{\vec p\,'\!\sigma'} \rangle 
=
\frac{\kappa^2}{4\pi} \frac{1}{2(2\pi)^3}\int d^3x \int d^3x' 
\left(e^{-i\vec p \cdot \vec x}
k_2(|\vec x-\vec x'|)
e^{i\vec p\, '\cdot \vec x'}
+  i \sigma \omega_p e^{-i\vec p \cdot \vec x}
k_1(|\vec x-\vec x'|)
\big(-i \sigma' \omega_{p'} e^{i\vec p\,' \cdot \vec x'}\big)
\right)
\nonumber
\\
&=\frac{\kappa^2}{4(2\pi)^4}  \int d^3x'  
\left\{
\int d^3x\,
k_2(|\vec x-\vec x'|)
e^{-i\vec p \cdot (\vec x -\vec x')}
+ \sigma \sigma'\omega_p \omega_{p'}
\int d^3x\,
k_1(|\vec x-\vec x'|)
e^{-i\vec p \cdot (\vec x -\vec x')}
\right\}
e^{-i(\vec p -\vec p \,')\cdot \vec x'}
\\
&=\frac{1}{2(2\pi)^3}\int d^3x'  
\left\{
(\omega_p)_+ +\sigma \sigma' \omega_{p'} (\omega_p)_+^0
\right\}
e^{-i(\vec p -\vec p \,')\cdot \vec x'}
=
\frac{1}{2} (1	+\sigma \sigma') (\omega_p)_+
\delta^{(3)}(\vec p - \vec p\,') =
\delta_{\sigma \sigma'}(\omega_p)_+\delta^{(3)}(\vec p - \vec p\,')
\,.
\nonumber
\end{align}
\end{widetext}
Again, $(\omega_p)_+$ and $(\omega_p)_+^0$ are zero if 
$|\vec p\,|<\kappa$. We have used that  $\int d^3 x \,
e^{-i \vec p \cdot \vec x } = (2\pi)^3\delta^{(3)}(\vec p\,)$ 
and that the integral involving the second kernel is $\int d^3x\,
k_1(|\vec x|) e^{-i\vec p \cdot \vec x } = \frac{4\pi}{\kappa^2}\frac{(\omega_p)_+^0}{\omega_p}$. 
The integral involving the first kernel is divergent. However, sense can be made of this integral with an adequate regularization. For instance, with dimensional regularization in dimension $d$, analytical continuation to $d=3$ provides $\int_{\mathbb{R}^3} d^3x\,k_2(|\vec x|) 
e^{-i\vec p \cdot \vec x } = \frac{4\pi}{\kappa^2}(\omega_p)_+$. 
That shows the positive-definiteness of the kernels of the proposed inner product.

With all that, we verify the unitary equivalence of configuration space and momentum space  representations: we see that the computation above is in agreement with \eqref{A15} for $|\vec p\,|\geq\kappa$, which is satisfied within the Hilbert space. This also confirms the fact that the inner product is non-degenerate only for functions whose Fourier transform have support on $|\vec p\,|>\kappa$, that is, for oscillatory solutions of \eqref{KGtacheq}, exactly like \eqref{KGtachinnpromom}. 

We should remark that using the usual scalar product \eqref{KGinnpro} would lead, on the contrary, to
$\langle \psi_{\vec p\sigma} , \psi_{\vec p\,'\!\sigma'} \rangle = 
\frac{1}{2}(\sigma + \sigma') \omega_p \delta^{(3)}(\vec p - \vec p\,')$, which is in clear conflict with the positive-definiteness of the inner product  \eqref{KGtachinnpromom} in momentum space representation and the restriction to the space of oscillatory solutions in configuration space representation.  

\begin{figure}[h!]
 \centering
   \includegraphics[width=8.6cm]{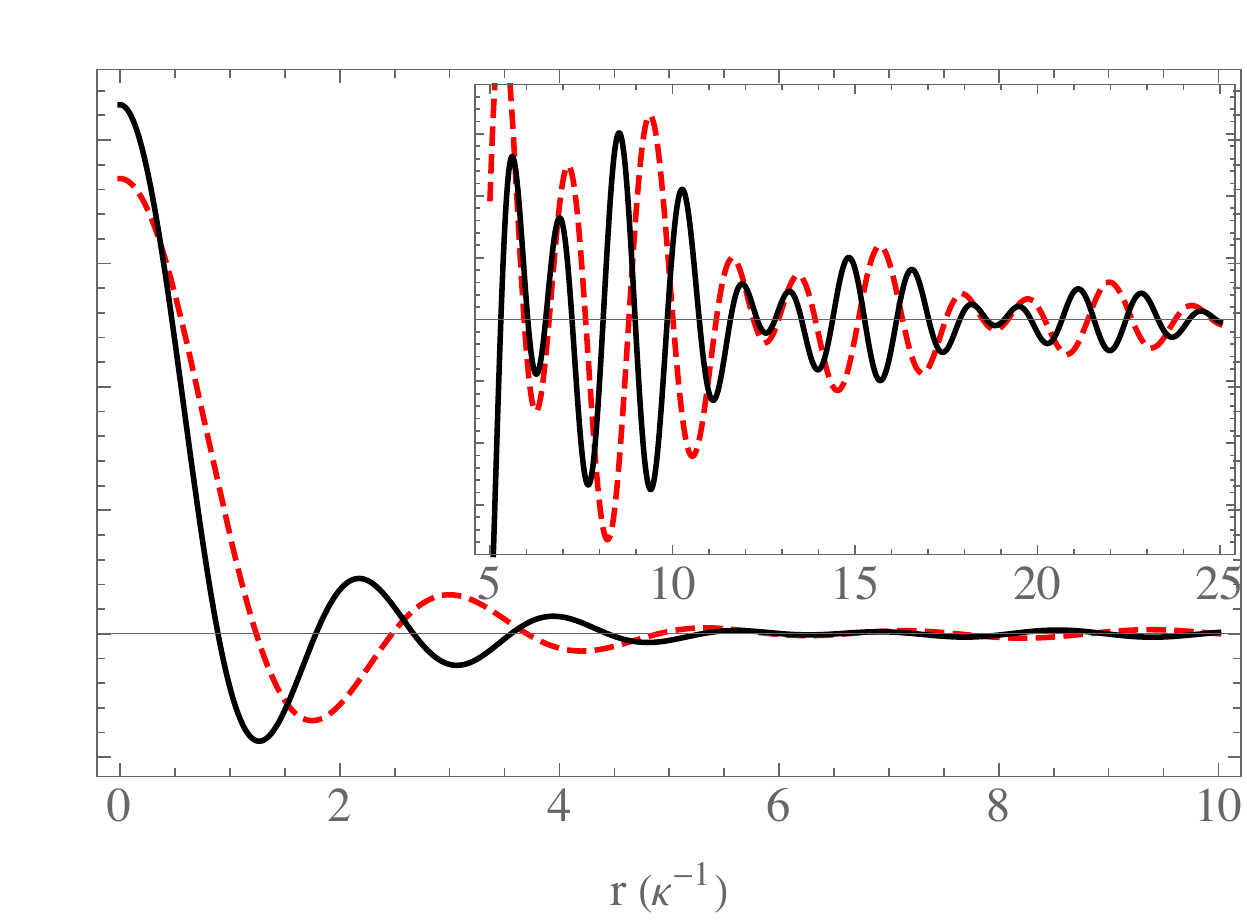}
   \caption{Configuration space radial wavefunctions $\varphi_{p_a p_b}(r)$ of two \textit{orthogonal} band-limited symmetric states with momentum space wavefunction $\varphi^\pm_{p_a p_b}(\vec p\,)= N \omega_p \theta(|\vec p\,|-p_a)\theta(p_b - |\vec p\,|)$. The radial coordinate $r$ is in units of $\kappa^{-1}$. The dashed red line represents the state with $p_a=2\kappa$, $p_b=3\kappa$ and the solid black represents $p_a=3\kappa$, $p_b=4\kappa$. A magnification over a wider domain is shown.}
   \label{fig1}
 \end{figure}

We can now work out an example. Consider the isotropic states defined in momentum space to have compact support: $\varphi^\pm_{p_a p_b}(\vec p\,)= N \omega_p \theta(|\vec p\,|-p_a)\theta(p_b - |\vec p\,|)$, where $\theta(x)$ is the Heaviside step function, $N$ a normalizing constant and $p_b -p_a$ is the bandwidth ($p_b > p_a > \kappa>0$). We take symmetric states with $\varphi^+_{p_a p_b}(\vec p\,)=\varphi^-_{p_a p_b}(\vec p\,)$ for simplicity, as this implies $\dot\varphi_{p_a p_b}(\vec x)=0$ (see \eqref{FTd}), whereas using \eqref{FT} we find
$\varphi_{p_a p_b}(\vec x)= \sqrt{2}N 
\Big(
\big(\frac{p_b}{|\vec x|}\big)^{\frac{3}{2}}J_{\frac{3}{2}}(p_b|\vec x|) 
-\big(\frac{p_a}{|\vec x|}\big)^{\frac{3}{2}}J_{\frac{3}{2}}(p_a|\vec x|)
\Big)$, where $J_n(x)$ is a Bessel function of the first kind. Two of such states with non-overlapping bandwidths are obviously orthogonal in view of \eqref{KGtachinnpromom}. In configuration space this is not obvious given that the inner product \eqref{KGtachinnpro} is non-local. See Fig.~\ref{fig1} for an illustration. The scalar product \eqref{KGinnpro} always yields $\langle\varphi_{p_a p_b} ,\varphi_{p_c p_d} \rangle_{st} =0$ for all those band-limited symmetric states and therefore does not reproduce the Hilbert space structure of $\mathcal H_m$, which is only possible due to the non-local nature of \eqref{KGtachinnpro}.

\textit{Projection on the Hilbert space.---}
Let us show another manifestation of the fact that initial conditions for \eqref{KGtacheq} can not be arbitrary functions on $\Sigma$ if the corresponding solutions $\phi(x)$ are to support a unitary and irreducible representation of $\mathcal P_+^\uparrow$ (we now deepen in some ideas already present in Ref.~\onlinecite{Feinberg1}). Inserting  \eqref{FT} and \eqref{FTd} in \eqref{IFT}, we obtain $(\omega_p)_+^0 \phi^\pm(\vec p\,)\equiv\tilde \phi^\pm(\vec p\,)$, that is, the identity as long as $ \phi^\pm(\vec p\,)$ has support on $|\vec p\,|\geq\kappa$ only. Now, if we use \eqref{IFT} in \eqref{FT}, we get:
\begin{align*}
&\frac{1}{2(2\pi)^3}\!\!
\int_{|\vec p\,|>\kappa}\!\!\frac{d^3p}{\omega_p}
	\bigg\{\!\int \!\!d^3x' \Big(\! (\omega_p)_+ \phi(\vec x')\!+\! 
	i (\omega_p)_+^0 \dot\phi(\vec x')\!\Big)
\\	
&+\int \!\!d^3x' \Big(\! (\omega_p)_+ \phi(\vec x')- 
	i (\omega_p)_+^0 \dot\phi(\vec x')\!\Big)\bigg\}e^{i\vec p \cdot (\vec x-\vec x\,')}
\\
&= 	\frac{1}{2(2\pi)^3}
\int \!\!d^3x' \bigg\{
\int_{|\vec p\,|>\kappa} \!\!\!\!d^3p \,2(\omega_p)_+^0 e^{i\vec p \cdot (\vec x-\vec x\,')}
\bigg\}\phi(\vec x')
\\
&=\phi(\vec x)-\int \!\!d^3x'\left(\frac{\kappa}{2\pi}\right)^{\frac{3}{2}}
\frac{J_{\frac{3}{2}}(\kappa|\vec x- \vec x'|)}{|\vec x- \vec x'|^{\frac{3}{2}}}\phi(\vec x')
\equiv \tilde\phi(\vec x) \,, 
\end{align*}  
where we have written the integral as $\int_{|\vec p\,|>\kappa} = \int_{\mathbb R^3}-\int_{|\vec p\,|<\kappa}$, the first term giving a Dirac delta and the second one the kernel $k_{3/2}(r)=\left(\frac{\kappa}{2\pi}\right)^{\frac{3}{2}}\frac{J_{\frac{3}{2}}(r)}{r^{3/2}}$, with a Bessel function of the first kind. An analogous result is obtained for $\dot\phi(\vec x)$. 

That may look puzzling at first sight: the expected result would be the identity. However, it must be noted that the convolution with the kernel $k_{3/2}(r)$ is zero for functions $\phi(\vec x)$ and $\dot\phi(\vec x)$ whose Fourier transforms $\hat\phi(\vec p\,)$ and $\hat{\dot \phi}(\vec p\,)$ have support on $|\vec p\,|\in [\kappa,+\infty )$. Therefore, the convolution with the kernel $\delta^{(3)}(\vec x - \vec x') - k_{3/2}(\kappa|\vec x - \vec x'|)$ \textit{projects} onto the Hilbert space $\mathcal H_{osc}$. As a consequence, within the Hilbert space, supporting the $m^2< 0$ irreducible representation of $\mathcal{P}_+^\uparrow$, the composition of \eqref{FT}-\eqref{IFT} is indeed the identity, as expected. 

A further comment is in order: the convolution operator with the kernel $\delta^{(3)}(\vec x - \vec x') - k_{3/2}(\kappa|\vec x - \vec x'|)$ is a non-trivial operator in $L^2(\mathbb R^3)$ which commutes with the operators of the Poincar\'e group. However, it becomes the identity when we restrict to the irreducible representation on $\mathcal H_{osc}$, a fact compatible with Schur's lemma.

\begin{figure}[h!]
 \centering
   \includegraphics[width=8.6cm]{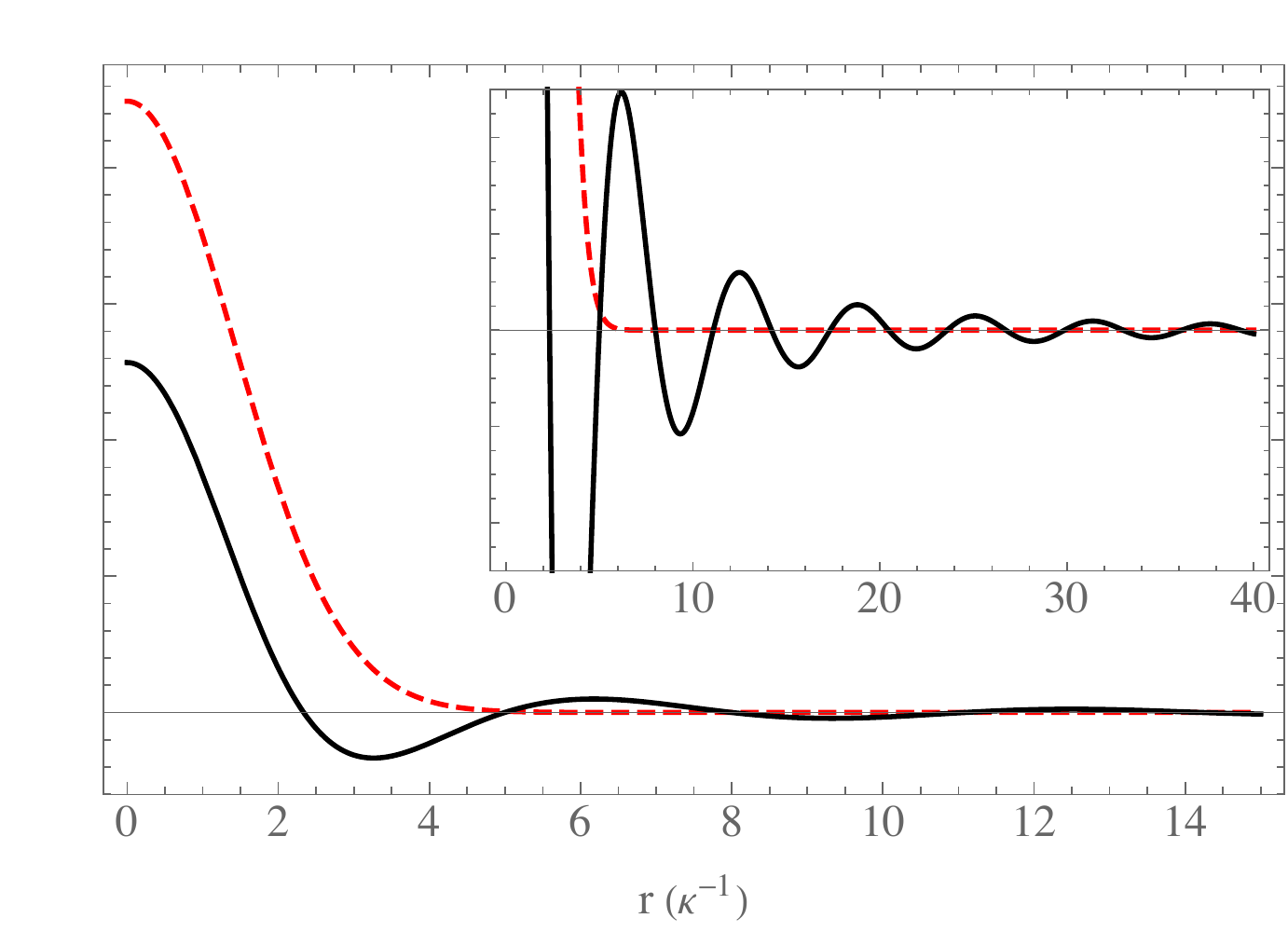}
   \caption{The non-admisible Gaussian wavefunction $\phi(r)$ as initial condition (dashed red) and its corresponding (admisible) projection $\tilde \phi(r)$ on the Hilbert space of oscillatory solutions for $\sigma=2/\kappa$ (solid black). The radial coordinate $r$ is in units of $\kappa^{-1}$. A magnification over a wider domain is shown to appreciate the oscillatory tail.}
   \label{fig2}
 \end{figure}

Let us check all that with an example. As initial conditions for \eqref{KGtacheq}, one might consider familiar Gaussian wavepackets. For simplicity, we focus on a spherically symmetric initial state $\phi(\vec x)=\frac{C}{(2\pi)^{3/2}\sigma^3} \exp({-\frac{|\vec x|^2}{2\sigma^2}})$ (equivalent results are obtained for Gaussian $\dot\phi(\vec x)$, although we are not interested in the subsequent time evolution). The result of the projection operation on the Hilbert space can be seen in Fig.~\ref{fig2}: we observe that the well-localized Gaussian wavefunction acquires an oscillatory tail that decays slower than the original one; actually, $\tilde\phi(\vec x)\thicksim \frac{\cos(\kappa|\vec x|)}{|\vec x|^2}$, while $\phi(\vec x)\thicksim  \exp({-\frac{|\vec x|^2}{2\sigma^2}})$. This turns out to be a general fact: no initial conditions for \eqref{KGtacheq} localized in space can lead to oscillatory solutions; any $\phi(\vec x)$ and $\dot \phi(\vec x)$ are necessarily spread over the entire Cauchy hypersurface $\Sigma$ \cite{Feinberg1}.

\textit{Discussion.---}
Once the missing scalar $m^2<0$ representation in configuration space have been realized, and the corresponding inner product established, there are still several issues to be addressed. 
Given that the inner product \eqref{KGtachinnpro} (as well as \eqref{KGtachinnprocov}) is positive definite, it seems possible to define a Relativistic Quantum Mechanics for wavefunctions satisfying \eqref{KGtacheq}.
  In fact, we have already seen that a suitable definition of the Hilbert space in terms of oscillatory, delocalized functions is necessary, as well as a non-local inner product. If possible, a precise definition of the density of probability in configuration space would also be required, a definition that might include a non-local operation such as the convolution with certain oscillatory kernels (for the case of the Klein-Gordon equation with $m^2>0$, see Ref. \onlinecite{Fonda}; see also Ref. \onlinecite{Mostafazadeh}). However, it must be pointed out that these circumstances are not that new: we have already encountered an analogous situation for the quantum mechanics in momentum space of a free particle moving on a sphere \cite{S3mom}, which turns out to be completely equivalent to the usual one in configuration space \cite{S3config}. Of course, whether the delocalization of the wavefunction implies that the field is physically delocalized in space, raising the question of the existence of a particle-like behavior, requires further study, as we have already pointed out that a probabilistic interpretation in accordance with the inner product \eqref{KGtachinnpro} is to be developed.  In any case, these peculiar features of the free theory may have implications in the analysis and even in the very definition of causality for an interacting `tachyon-like' quantum field theory.

Besides the already mentioned probabilistic interpretation, the non-scalar, infinite-dimensional representations \cite{Moses} remain to be found in configuration space. To determine whether scalar or non-scalar tachyon-like representation are to bear some physical significance at the fundamental level and/or fully develop a quantum field theory, a careful analysis of causality is required in light of the delocalization property. For that aim, the form of those interactions compatible with Poincar\'e symmetry should also be analyzed, as well as the time-evolution. 

Finally, the framework described here may play some role as a piece in a broader physical theory. There have been many contributions in that direction, such as in string theory \cite{Sen,Armoni}, supersymmetry \cite{ellis}, quantum field theory in curved spacetime \cite{Landulfo}, cosmology \cite{Paddy,Teixeira}, spontaneous symmetry breaking \cite{Felder1,Felder2}, QCD \cite{Yokota}, condensed matter \cite{Oriekhov} or even in the study of liquids \cite{Trachenko}. Maybe the analysis of quantum effects in those contexts might require a revision with the new inner product \eqref{KGtachinnpro} (or a generalization) in mind. 
In fact, we want to stress that the aim of this Letter is more preliminary and/or fundamental: we have just provided the inner product \eqref{KGtachinnprocov}-\eqref{KGtachinnpro} and, from that, the new configuration space realization of the scalar $m^2<0$ representation of $\mathcal P_+^\uparrow$.



\bigskip

The authors are very grateful to E.R. Arriola for fruitful discussions. We thank the Spanish Ministerio de Ciencia e Innovaci\'on (MICINN) for financial support (FIS2017-84440-C2-2-P) and  V.A. acknowledges financial support from the State Agency for Research of the Spanish MCIU through the `Center of Excellence Severo Ochoa' award for the Instituto de Astrof\'\i sica de Andaluc\'\i a (SEV-2017-0709). J.G. acknowledges financial support from the Spanish MICINN (PGC2018-097831-B-I00).


%
%

%



\end{document}